\documentclass[%
aps, %
prb, 
twocolumn, 10pt, superscriptaddress, 
a4paper, %
amsfonts, amsmath, amssymb]{revtex4-2}
\usepackage[sort&compress]{natbib}
\usepackage{newtxtext}
\usepackage[varg]{newtxmath}
\usepackage{ascmac}
\usepackage{mathrsfs}
\usepackage{bm}
\usepackage{color}
\usepackage{graphicx}
\usepackage{bm}
\usepackage{braket}
\usepackage{mathrsfs}
\usepackage[normalem]{ulem}
\usepackage{hhline}
\usepackage{hyperref}
\hypersetup{%
	unicode=false, %
	bookmarksnumbered=true, %
	bookmarkstype=toc, %
	breaklinks=true, %
	colorlinks=true, %
	pdfborder={0 0 1}, %
	linkcolor=blue, %
	citecolor=blue, %
	urlcolor=blue, %
	pdftoolbar=true, %
	pdfmenubar=true, %
	pdffitwindow=false, %
	pdfstartview={FitH} %
}
\newcommand{\kB}{k_{\mathrm{B}}}

\newcommand{\eF}{\epsilon_{\mathrm{F}}}
\newcommand{\br}{\bm{r}}
\newcommand{\bk}{\bm{k}}
\newcommand{\bp}{\bm{p}}

\newcommand{\zi}{i}
\renewcommand{\Im}{\mathrm{Im}\,}
\newcommand{\tr}{\mathrm{tr}\,}

\newcommand{\R}{\mathrm{R}}
\newcommand{\A}{\mathrm{A}}
\newcommand{\X}{\mathrm{X}}
\newcommand{\Y}{\mathrm{Y}}
\newcommand{\re}{\operatorname{\mathbb{R}e}}
\newcommand{\im}{\operatorname{\mathbb{I}m}}
\begin{document}
%
\title{Seebeck effect of Dirac electrons}
\date{\today}
\author{Junji Fujimoto}
\email[E-mail address: ]{fujimoto@hosi.phys.s.u-tokyo.ac.jp}
\affiliation{Department of Physics, University of Tokyo, Bunkyo, Tokyo 113-0033, Japan}
\author{Masao Ogata}
\affiliation{Department of Physics, University of Tokyo, Bunkyo, Tokyo 113-0033, Japan}
\affiliation{Trans-scale Quantum Science Institute, University of Tokyo, Bunkyo-ku, Tokyo 113-0033, Japan}

\begin{abstract}
We study the Seebeck effect in the three-dimensional Dirac electron system based on the linear response theory with Luttinger's gravitational potential.
The Seebeck coefficient $S$ is defined by $S = L_{12} / L_{11} T$, where $T$ is the temperature, and $L_{11}$ and $L_{12}$ are the longitudinal response coefficients of the charge current to the electric field and to the temperature gradient, respectively; $L_{11}$ is the electric conductivity and $L_{12}$ is the thermo-electric conductivity.
We consider randomly-distributed impurity potentials as the source of the momentum relaxation of electrons and microscopically calculate the relaxation rate and the vertex corrections of $L_{11}$ and $L_{12}$ due to the impurities.
It is confirmed that $L_{11}$ and $L_{12}$ are related through Mott's formula in low temperatures when the chemical potential lies above the gap ($|\mu| > \Delta$), irrespective of the linear dispersion of the Dirac electrons and unconventional energy dependence of the lifetime of electrons.
On the other hand, when the chemical potential lies in the band gap ($|\mu| < \Delta$), Seebeck coefficient behaves just as in conventional semiconductors: Its dependences on the chemical potential $\mu$ and the temperature $T$ are partially captured by $S \propto (\Delta - \mu) / \kB T$ for $\mu > 0$.
The Seebeck coefficient takes the relatively large value $|S| \simeq 1.7 \,\mathrm{m V/K}$ at $T \simeq 8.7\,\mathrm{K}$ for $\Delta = 15 \,\mathrm{m eV}$ by assuming doped bismuth.\end{abstract}
\maketitle

\section{Introduction}
Heat flow is accompanied by electric current in metallic materials.
This phenomenon can be understood since one of the heat carriers in materials are electrons or holes, which have charge.
The effect is called Seebeck effect, which is one of the thermoelectric effects.
When a material is in an open circuit, electric voltage arises in the longitudinal direction to the temperature gradient.
This thermo-electomotive force $\bm{E}_{\mathrm{emf}}$ is characterized by the Seebeck coefficient $S$ as $\bm{E}_{\mathrm{emf}} = S \bm{\nabla} T$, where $\bm{\nabla} T$ is the temperature gradient.
It is importants to find materials having large Seebeck coefficient for harvesting waste heat and converting it into useful electrical power.

\begin{figure}[bt]
\centering
\includegraphics[width=\linewidth, clip]{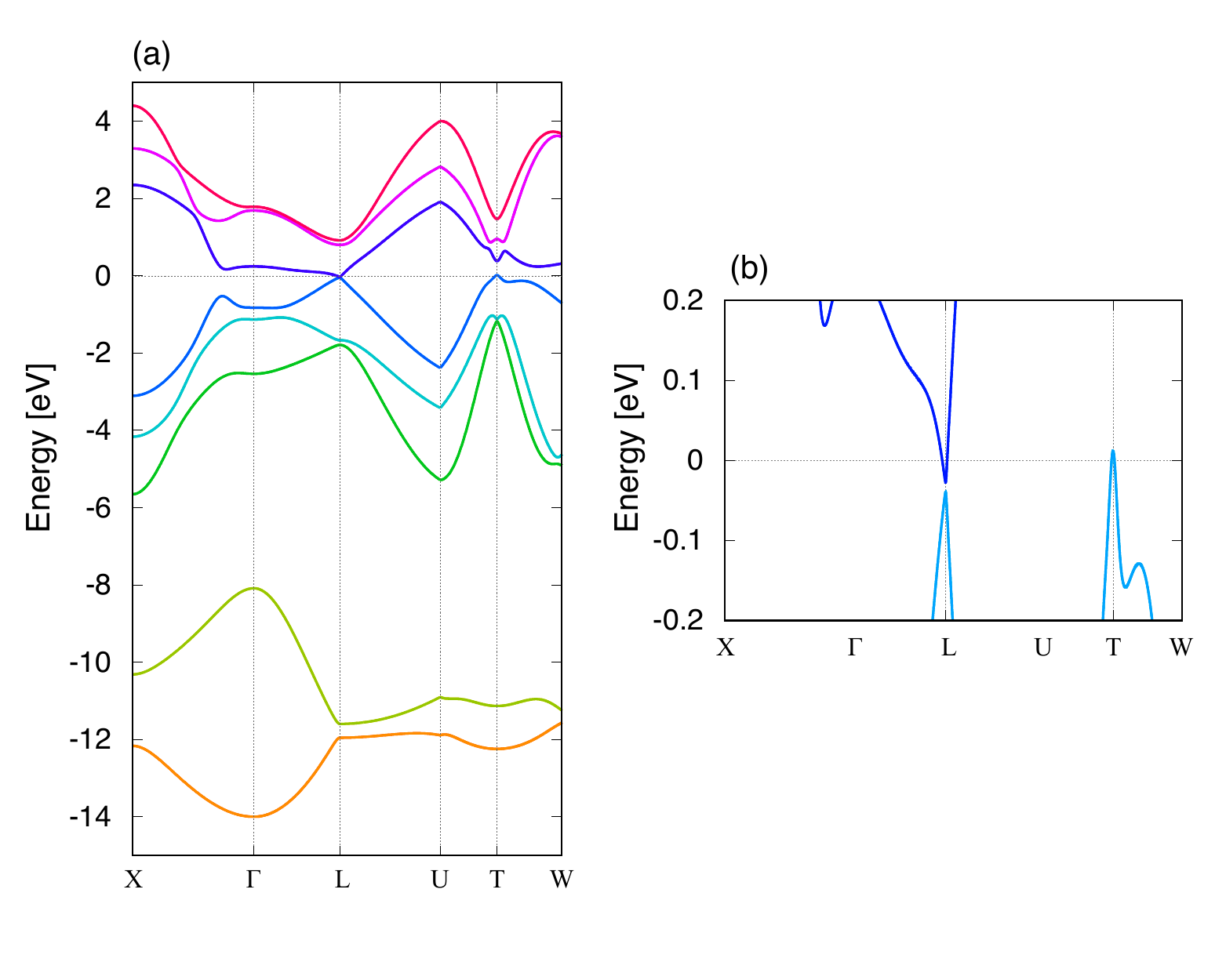}
\caption{\label{fig:dispersion}Band structure of bismuth along some symmetry lines (a)~for whole the energy scale and (b)~near the Fermi level.
The band structure is based on the tight-binding Hamiltonian given in Ref.~\onlinecite{liu1995}, and the symmetry points are given in Ref.~\onlinecite{falicov1965}.}
\end{figure}

Bismuth is the material in which the Seebeck effect was observed for the first time~\cite{seebeck1822}.
Despite its long history, the microscopic calculation of the Seebeck effect in bismuth has not been done yet.
From the current understanding for the bismuth crystal, there are electron and hole pockets in the $L$ and $T$ points in the Brillouin zone, respectively, and electrons near the $L$ point can be described by a three-dimensional~(3D) Dirac Hamiltonian~\cite{cohen1960,wolff1964,fuseya2012}~(see Fig.~\ref{fig:dispersion}), in which the energy dispersion is linear with respect to the momentum measured from the $L$ point.
Thus, it is important to study whether this kind of linear dispersion leads to unusual temperature- or chemical-potential-dependences of the Seebeck coefficient compared with usual metals with quadratic momentum dispersion.
Furthermore, it is well known that the impurity scattering in Dirac systems, e.g., in graphene, causes unconventional energy dependence of the relaxation rate~\cite{shon1998}.
Therefore, it is necessary to develop a microscopic theory on the relaxation rate for the massive 3D Dirac electrons using the self energy of the Green's functions as well as the vertex corrections in calculating the electronic conductivity and Seebeck coefficients.

In this paper, we study the Seebeck effect in the 3D Dirac electron system.
The Seebeck coefficient is defined using longitudinal response coefficients of the charge current to the electric field and to the temperature gradient, which correspond to the electric conductivity and thermo-electric conductivity~\cite{behnia2015,ogata2019}, respectively.
We calculate the response coefficients based on the linear response theory with Luttinger's gravitational potential~\cite{luttinger1964}.
We consider randomly-distributed impurity potentials as the source of the momentum relaxation of electrons, and the relaxation rate and the vertex corrections of $L_{11}$ and $L_{12}$ due to the impurities are microscopically calculated, which shows different energy dependences from the quadratic momentum dispersion.
We will confirm that the two response coefficients are related through Mott's formula in low temperatures.
For the chemical potential dependence of the Seebeck coefficient $S$, we find that $S$ has a peak structure when the chemical potential lies in the band gap at sufficiently low temperatures, while $S$ has monotonic behavior when the system is metallic.
We also find a similar peak structure for the temperature dependence of $S$, as found in the chemical potential dependence.
We show that these peak structures are partially understood by the phenomenological theory used in semiconductors, which suggests the Seebeck coefficient is proportional to $(\Delta - \mu) / \kB T$ for $0 < \mu < \Delta$, where $\Delta$ is the band gap, $\mu$ is the chemical potential, $\kB$ is the Boltzmann constant, and $T$ is the temperature.
This behavior of the Seebeck coefficient indicates that the thermoelectric property of the 3D Dirac electrons in the band gap is same as that of conventional semiconductors.

By doping to bismuth, the Fermi level may be tuned without changing the band structure.
Appropriate doping leads to the chemical potential inside the band gap in the $L$ point, where the Seebeck coefficient has the peak structure in the temperature dependence as mentioned above.
We evaluate the Seebeck coefficient near the peak and find the value $|S| \simeq 1.7 \,\mathrm{m V/K}$ at $T \simeq 8.7\,\mathrm{K}$ for $\Delta = 15 \,\mathrm{m eV}$.
Although the contribution to the Seebeck effect from the holes near the $T$ point should be taken into account, it would not be significant because it should be proportional to the temperature at low temperatures.

Here, we mention the method of Luttinger's gravitational potential.
In the linear response theory, the Kubo formula is formulated based on external mechanical forces.
A mechanical force $F$ couples to the physical quantity $\hat{A}$ through its Hamiltonian; $H_{\mathrm{ext}} = \hat{A} F$.
The Kubo formula indicates that the response function of a physical quantity $\hat{B}$ to the external force $F$ is given by the correlation function between $\hat{B}$ and $\hat{A}$.
On the other hand, the temperature gradient is a statistical force, which cannot be written by Hamiltonian.
It is not trivial whether the Kubo formula is valid for responses to a statistical force.
Luttinger introduced a fictional gravitational potential, which is a mechanical force and couples to the Hamiltonian density~\cite{luttinger1964}.
Note that the gradient of the fictional potential couples the thermal current.
Then, one can apply the Kubo formula to the gravitational potential, and from the fact that the responses to the mechanical force and to the statistical force are equivalent for the nonequilibrium components, which is called Einstein's relation, one obtain the response to the temperature gradient.

\section{Model and Green's function}
Following Ref.~\onlinecite{fuseya2012}, we consider the effective (isotropic) Dirac Hamiltonian,
\begin{align}
\mathcal{H}_{\mathrm{D}}
	& =
	\begin{pmatrix}
		\Delta
	&	i \hbar v \bm{k} \cdot \bm{\sigma}
	\\	- i \hbar v \bm{k} \cdot \bm{\sigma}
	&	- \Delta
	\end{pmatrix}
	= - \hbar v \rho_2 \bm{k} \cdot \bm{\sigma} + \Delta \rho_3
\label{eq:Dirac_Hamiltonian}
,\end{align}
where $v$ is the velocity, $\bm{\sigma} = (\sigma^x, \sigma^y, \sigma^z)$ is the Pauli matrix in spin space, and $\rho_i$ with $i = 1,2,3$ represents the Pauli matrix in particle-hole space.
We use $\rho_0$ and $\sigma^0$ as the unit matrices when emphasizing them.
We also consider the point-like impurity potential $V = u \sum_i \rho_0 \sigma^0 \delta (\br - \bm{R}_i)$, where $u$ is the potential strength, and $\bm{R}_i$ represents the positions of the impurities.
The total Hamiltonian is given by $\mathcal{H} = \mathcal{H}_{\mathrm{D}} + V$.

By taking the average on the positions of impurities~\cite{kohn1957,mahan2000}, the retarded self energy in the Born approximation is given by
\begin{align}
\varSigma^{\R} (\epsilon)
	& = \frac{n_{\mathrm{i}} u^2}{\Omega} \sum_{\bm{k}} G^{(0)}_{\bm{k}} (\epsilon + i 0)
\label{eq:retarded_self_energy}
,\end{align}
where $n_{\mathrm{i}}$ is the impurity concentration (we assume $n_{\mathrm{i}} \ll 1$), $\Omega$ is the system volume, and the bare Green's function is defined by $G^{(0)}_{\bm{k}} (\epsilon + i 0) = (\epsilon + \mu - \mathcal{H}_{\rm D} + i 0)^{-1}$.
 The imaginary part is evaluated as
\begin{align}
\Im \varSigma^{\R} (\epsilon)
	& = - \gamma_0 (\epsilon+\mu) \rho_0 - \gamma_3 (\epsilon+\mu) \rho_3
\label{eq:Im_self-energy}
\end{align}
with the damping constants
\begin{subequations}
\begin{align}
\gamma_0 (\epsilon)
	& = \frac{\pi}{2} n_{\rm i} u^2 \nu (\epsilon)
, \\
\gamma_3 (\epsilon)
	& = \frac{\pi}{2} n_{\rm i} u^2 \frac{\Delta}{\epsilon} \nu (\epsilon)
,\end{align}
\label{eq:damping-constants}%
\end{subequations}
where $\nu (\epsilon)$ is the density of states~(DOS),
\begin{align}
\nu (\epsilon)
	& = \frac{1}{\Omega} \sum_{\bm{k}, \eta = \pm} \delta (\epsilon - \eta \varepsilon_k)
\notag	\\	&
	= \frac{|\epsilon|}{2 \pi^2 \hbar^3 v^3} \sqrt{\epsilon^2 - \Delta^2} \sum_{\eta} \Theta ( \eta \epsilon - \Delta )
\label{eq:DOS}
\end{align}
with $\Theta (x)$ being the Heaviside step function.

We note that it is not obvious how the impurity potential is expressed in the basis of the Dirac Hamiltonian, as noted in Ref.~\onlinecite{fukazawa2017}.
For simplicity, the impurity potential is here assumed to be proportional to $\rho_0 \sigma^0$.
We also note that bismuth has three $L$ points in Brillouin zone, and the inter-valley scatterings for the point-like impurity potential should be considered~\cite{shon1998}.
However, we neglect the inter-valley scatterings at the first step.

Then, the retarded Green's function of this system is given as
\begin{align}
G^{\mathrm{R}}_{\bm{k}} (\epsilon)
	& = \frac{1}{ D^{\mathrm{R}}_{\bm{k}} (\epsilon+\mu) }
		\Bigl( g^{\mathrm{R}}_0 (\epsilon+\mu) + \rho_2 \bm{g}^{\mathrm{R}}_2 (\bm{k}) \cdot \bm{\sigma} + \rho_3 g^{\mathrm{R}}_3 (\epsilon+\mu) \Bigr)
\label{eq:retarded_G_def}
\end{align}
with
\begin{subequations}
\begin{align}
D^{\mathrm{R}}_{\bm{k}} (\epsilon)
	& = ( \epsilon + i \gamma_0 (\epsilon) )^2
	- \hbar^2 v^2 k^2
	- (\Delta - i \gamma_3 (\epsilon))^2
\label{eq:DR_k}
, \\
g^{\mathrm{R}}_0 (\epsilon)
	& = \epsilon + i \gamma_0 (\epsilon)
, \\
\bm{g}^{\mathrm{R}}_2 (\bm{k})
	& = - \hbar v \bm{k}
, \\
g^{\mathrm{R}}_3 (\epsilon)
	& = \Delta - i \gamma_3 (\epsilon)
.\end{align}
\end{subequations}

The denominator can be written as
\begin{align}
D^{\R}_{\bk} (\epsilon)
	& = \prod_{\eta = \pm} ( \epsilon - \eta \epsilon_k + \zi \Gamma (\epsilon))
,\end{align}
where the eigenenergies are $\pm \varepsilon_k = \pm \sqrt{\hbar^2 v^2 k^2 + \Delta^2}$ and $\Gamma (\epsilon)$ represents the damping of the electron
\begin{align}
\Gamma (\epsilon)
	= \frac{\pi}{2} n_{\mathrm{i}} u^2 \left( 1 + \frac{\Delta^2}{\epsilon^2} \right) \nu (\epsilon)
\label{eq:tau}
,\end{align}
where we have used the on-shell condition $\epsilon = \eta \epsilon_k$.

\begin{figure}[htpb]
\centering
\includegraphics[width=0.9\linewidth]{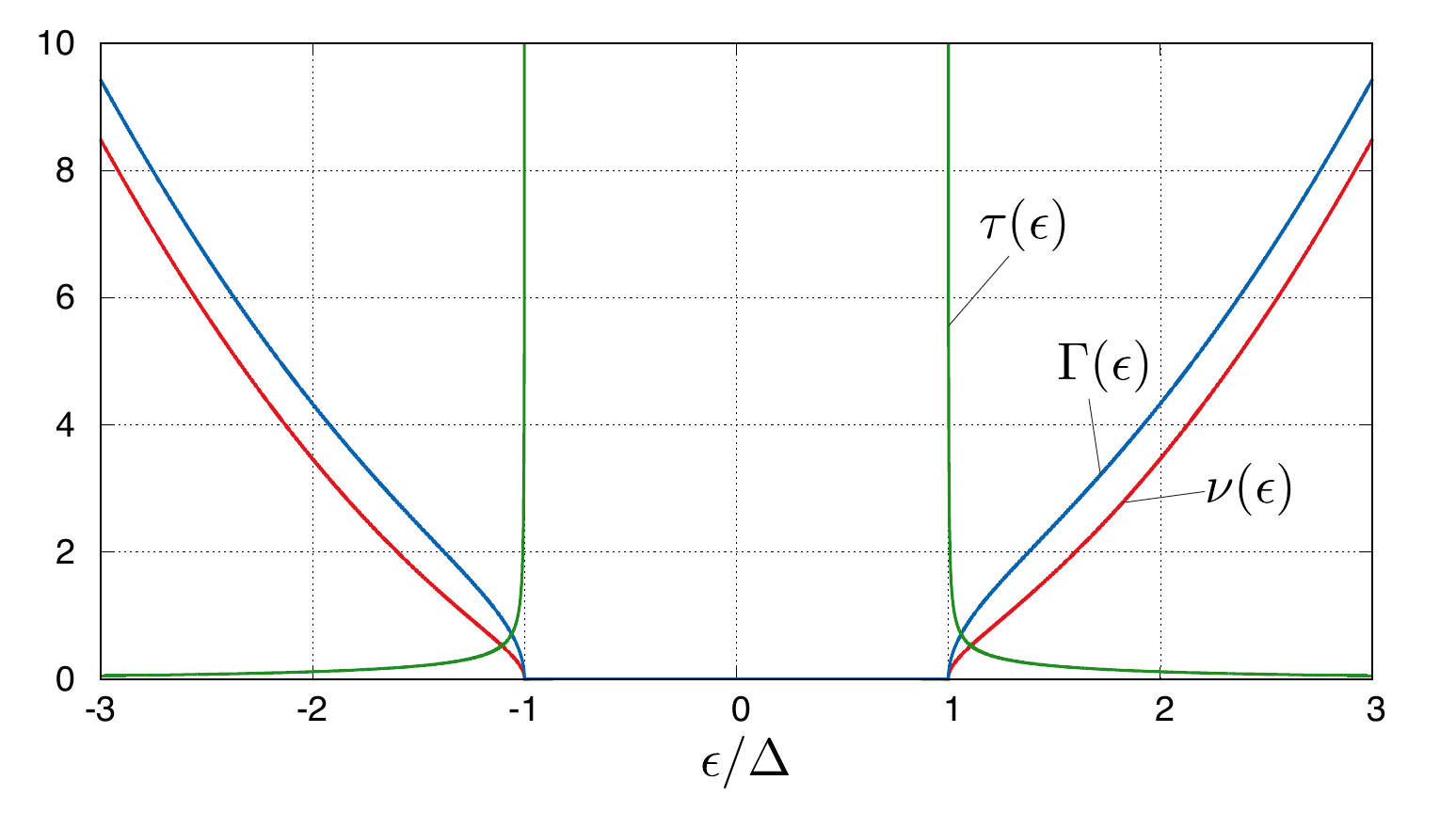}
\caption{\label{fig:damping}Energy dependences of DOS $\nu (\epsilon)$, the damping $\Gamma (\epsilon)$, and the lifetime $\tau (\epsilon)$ in the Born approximation.}
\end{figure}

Here we give some comments on the damping constants $\gamma_0 (\epsilon)$, $\gamma_3 (\epsilon)$ and the damping $\Gamma (\epsilon)$ and on their energy dependences.
For the parabolic dispersion in the absence of magnetization, there is one kind of damping constant, that is the $\gamma_0$-type.
The two-dimensional Dirac electron system without the mass gap~\cite{shon1998} has also only the $\gamma_0$-type damping constant.
However, the 3D Dirac electron system with mass gap has the two kinds of damping constants as shown in Eq.~(\ref{eq:Im_self-energy}).
This is not because of 3D, but because the system has the mass gap.
Generally speaking, the damping constants in the Born approximation depends on the physical quantities in the Hamiltonian, such as energy, mass gap, and magnetization~\cite{fujimoto2014,fujimoto2018}.
Accordingly, the damping of the electron $\Gamma (\epsilon)$ has a different dependence on the energy from DOS as shown in Fig.~\ref{fig:damping}.
Here the lifetime is defined by $\hbar / 2 \tau (\epsilon) = \Gamma (\epsilon)$.

The velocity operator is given by
\begin{align}
v_i
	& = \frac{1}{\hbar} \frac{\partial \mathcal{H}_{\mathrm{D}}}{\partial k_i}
	= - v \rho_2 \sigma^i
\end{align}
and the electric current operator in the second quantization is obtained as
\begin{align}
J_i
	& = e v \sum_{\bk} c^{\dagger}_{\bk} \rho_2 \sigma^i c^{}_{\bk}
,\end{align}
where $c_{\bk}^{(\dagger)}$ is the Fourier component of the field operator, and $- e$ is the electron charge.
The thermal current operator in the imaginary time domain is calculated as~\cite{jonson1990,kontani2003a,ogata2019}
\begin{align}
\bm{J}_{\mathrm{Q}}
	= - \frac{v}{2} \sum_{\bk} \left(
		\dot{c}^{\dagger}_{\bk} \rho_2 \bm{\sigma} c^{}_{\bk}
		- c^{\dagger}_{\bk} \rho_2 \bm{\sigma} \dot{c}^{}_{\bk}
	\right)
,\end{align}
where $\dot{c}_{\bk}^{(\dagger)} = d c_{\bk}^{(\dagger)} / d \tau$ is the imaginary time derivative (see Appendix~\ref{apx:heat_current}).

\section{Seebeck effect}
In this paper, we consider the longitudinal charge current response to the electric field $E_x$ and to the temperature gradient $- \nabla_x T / T$, which is shown as
\begin{align}
\langle J_x \rangle
	& = L_{11} E_x + L_{12} \left( - \frac{\nabla_x T}{T} \right)
\end{align}
with the response coefficients $L_{11}$ and $L_{12}$.
The Seebeck coefficient $S$ is given as~\cite{mahan2000}
\begin{align}
S
	& = \frac{L_{12}}{L_{11} T}
\label{eq:Seebeck}
.\end{align}

From the linear response theory, the response coefficients are calculated from
\begin{subequations}
\begin{align}
L_{11}
	& = \lim_{\omega \to 0} \frac{\chi_{c} (\omega) - \chi_{c} (0)}{\zi \omega}
, \\
L_{12}
	& = \lim_{\omega \to 0} \frac{\chi_{Q} (\omega) - \chi_{Q} (0)}{\zi \omega}
,\end{align}%
\end{subequations}
where $\chi_{i}$ with $i = c, Q$ is evaluated from the corresponding thermal correlation functions
\begin{subequations}
\begin{align}
\chi_c (\zi \omega_{\lambda})
	& = \frac{1}{\Omega} \int_0^{\beta} \mathrm{d}\tau e^{\zi \omega_{\lambda} \tau}
		\langle \mathrm{T}_{\tau} J_x (\tau) J_x \rangle
\label{eq:chi_c}%
, \\
\chi_Q (\zi \omega_{\lambda})
	& = \frac{1}{\Omega} \int_0^{\beta} \mathrm{d}\tau e^{\zi \omega_{\lambda} \tau}
		\langle \mathrm{T}_{\tau} J_x (\tau) J_{\mathrm{Q}, x} \rangle
\label{eq:chi_Q}%
,\end{align}
\label{eqs:chi}%
\end{subequations}
by taking the analytic continuation $\zi \omega_{\lambda} \to \hbar \omega + \zi 0$.
Here, $\beta = 1/\kB T$ is the inverse temperature, $\omega_{\lambda} = 2 \pi \lambda \kB T$ with an integer $\lambda$ is the Matsubara frequency of bosons, $\mathrm{T}_{\tau}$ is the imaginary time ordering operator, and $J_x (\tau) = e^{\tau \mathcal{H}} J_x e^{- \tau \mathcal{H}}$ is the Heisenberg representation in the imaginary time.

To be exact, we first calculate the charge current response to Luttinger's gravitational potential based on the linear response theory, and then use the Einstein's relation, which leads to the above formulation~\cite{luttinger1964,smrcka1977,cooper1997,Kohno2014,ogata2019}.
Note that we need no care to the local equilibrium correction for the longitudinal component~\cite{smrcka1977,cooper1997,Kohno2014}, since it is obviously zero.

Rewriting Eqs.~(\ref{eqs:chi}) using thermal Green's function $\mathcal{G}_{\bk} (\tau) = - \langle \mathrm{T}_{\tau} c_{\bk} (\tau) c^{\dagger}_{\bk} \rangle$ with the following relation (see Appendix~\ref{apx:relation} for the derivation)
\begin{subequations}
\begin{align}
\langle \mathrm{T}_{\tau} c_{\bk} (\tau) \dot{c}^{\dagger}_{\bk} \rangle
	& = \frac{\mathrm{d}}{\mathrm{d}\tau} \mathcal{G}_{\bk} (\tau) + \delta (\tau)
\label{eq:derivatives-a}%
, \\
\langle \mathrm{T}_{\tau} \dot{c}_{\bk} c^{\dagger}_{\bk} (\tau) \rangle
	& = \frac{\mathrm{d}}{\mathrm{d}\tau} \mathcal{G}_{\bk} (-\tau) - \delta (\tau)
\label{eq:derivatives-b}%
,\end{align}
\label{eqs:derivatives}%
\end{subequations}
with $\dot{c}_{\bk}^{(\dagger)} = d c_{\bk}^{(\dagger)} / d \tau$, and then taking the analytic continuation $\zi \omega_{\lambda} \to \hbar \omega + \zi 0$, we obtain
\begin{align}
L_{11}
	& = \int_{-\infty}^{\infty} \mathrm{d}\epsilon
			\left( - \frac{\partial f}{\partial \epsilon} \right) \sigma (\epsilon + \mu)
\label{eq:L_11-tochu}
, \\
L_{12}
	& = \frac{1}{- e} \int_{-\infty}^{\infty} \mathrm{d}\epsilon
			\left( - \frac{\partial f}{\partial \epsilon} \right) \epsilon \sigma (\epsilon + \mu)
\label{eq:L_12-tochu}
,\end{align}
where $f (\epsilon) = ( e^{\beta \epsilon} + 1 )^{-1}$, and
\begin{align}
\sigma (\epsilon + \mu)
	& = \frac{\hbar e^2 v^2}{4 \pi \Omega}
	\sum_{\bk} \tr \left[
		\rho_2 \sigma^x G^{\R}_{\bk} (\epsilon) \Lambda_{2, x} G^{\A}_{\bk} (\epsilon)
	\right]
\label{eq:sigma-tochu}
.\end{align}
The derivations of Eqs.~(\ref{eq:L_11-tochu}) and (\ref{eq:L_12-tochu}) with Eq.~(\ref{eq:sigma-tochu}) are given in Appendix~\ref{apx:detail:correlation}.
The vertex $\Lambda_{2, x}$ is the full velocity vertex including the ladder-type vertex corrections,
\begin{align}
\Lambda_{2, x}
	& = \rho_2 \sigma^x
		+ \frac{n_{\mathrm{i}} u^2}{\Omega} \sum_{\bk} G^{\R}_{\bk} (\epsilon) \Lambda_{2, x} G^{\A}_{\bk} (\epsilon)
\label{eq:Lambda_2x}
,\end{align}
which can be solved as
\begin{align}
\Lambda_{2, x}
	= \frac{1}{1 - U} \rho_2 \sigma^x
	+ \mathcal{O} (n_{\mathrm{i}})
\label{eq:Lambda_2x_end}
\end{align}
with $U = U (\epsilon + \mu)$ and
\begin{align}
U (\epsilon)
	= \frac{1}{3} \frac{\epsilon^2 - \Delta^2}{\epsilon^2 + \Delta^2}
\end{align}
as in the previous work~\cite{fukazawa2017}.
The detailed calculation from Eq.~(\ref{eq:Lambda_2x}) to (\ref{eq:Lambda_2x_end}) is shown in Appendix~\ref{apx:detail:vc}.
Note that the vertex corrections have $\mathcal{O} (n_{\mathrm{i}}^0)$-contributions, since $\sum_{\bk} G^{\R}_{\bk} (\epsilon) \rho_{2} \sigma^x G^{\A}_{\bk} (\epsilon) \propto \tau \propto 1 / n_{\mathrm{i}} u^2$.
We then obtain
\begin{align}
\sigma (\epsilon + \mu)
	& = \frac{\hbar e^2 v^2}{\pi} \frac{1}{4 n_{\mathrm{i}} u^2} \tr \left[
		\rho_2 \sigma^x
		\left( \frac{n_{\mathrm{i}} u^2}{\Omega} \sum_{\bk} G^{\R}_{\bk} \Lambda_{2, x} G^{\A}_{\bk} \right)
	\right]
\notag \\
	& = \frac{\hbar e^2 v^2}{\pi} \frac{1}{4 n_{\mathrm{i}} u^2} \tr \left[
		\rho_2 \sigma^x \left( \Lambda_{2, x} - \rho_2 \sigma^x \right)
	\right]
\notag \\
	& = \frac{\hbar e^2 v^2}{\pi} \frac{1}{n_{\mathrm{i}} u^2} \frac{U}{1 - U}
.\end{align}
Here, we note that we calculated only the leading order with respect to $\mu \tau / \hbar \gg 1$ in $\sigma (\epsilon)$.
When one is to calculate the higher order contributions, the terms which include only $G^{\R}_{\bk}$ or $G^{\A}_{\bk}$ should also be calculated (see Appendix~\ref{apx:detail:correlation}).

\section{Results and Discussion}
\begin{figure*}[htpb]
\centering
\includegraphics[width=0.9\linewidth]{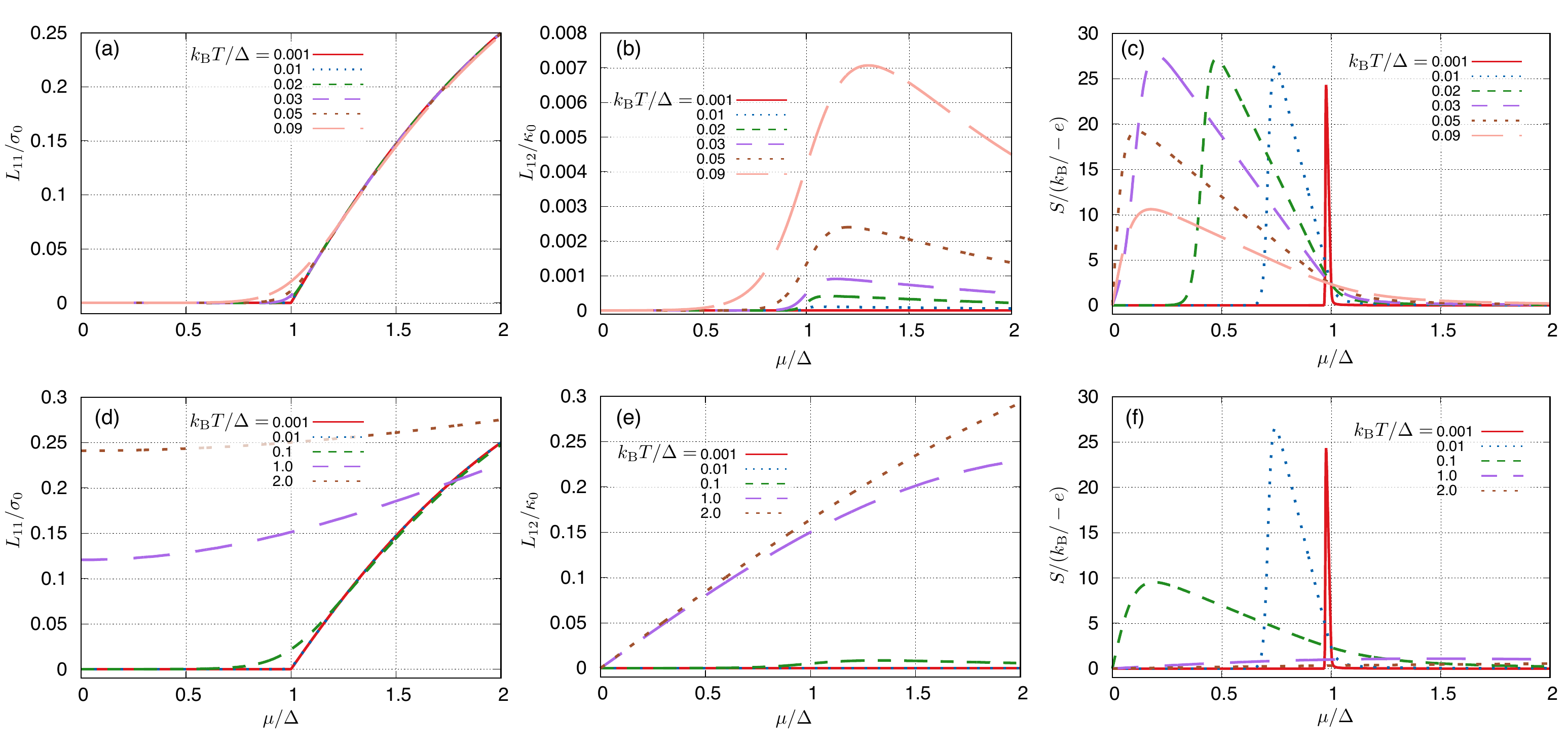}
\caption{\label{fig:vs_mu}%
Chemical potential dependences of the response coefficients $L_{11}$, $L_{12}$, and the Seebeck coefficient $S$.
(a)-(c): $L_{11}$, $L_{12}$ and $S$ at relatively low temperatures, (d)-(f): those in the wider range.
The units are given as $\sigma_0 = e^2 v^2 \tau_0 \nu_0$ and $\kappa_0 = - \sigma_0 \Delta / e$ with $\tau_0 = \hbar / \pi n_{\mathrm{i}} u^2 \nu_0$ and $\nu_0 = \Delta^2 / 2 \pi^2 \hbar^3 v^3$.
The difference between the temperature dependences of $L_{11}$ and $L_{12}$ leads to a peak structure of $S$ in the band gap at low temperatures, while such peak structures disappear above $\kB T / \Delta \gtrsim 1$.
The Seebeck coefficient takes $S \simeq - 20 \,\kB / e \simeq - 1.7 \, \mathrm{m eV / K}$ at $\kB T / \Delta = 0.05$ for $\Delta = 15 \,\mathrm{m e V}$, which corresponds to $T = 8.7 \,\mathrm{K}$.
}
\end{figure*}

We here summarize the result.
The longitudinal charge current response coefficient to the electric field, i.e. the electric conductivity, is calculated as
\begin{align}
L_{11}
	& = \int_{-\infty}^{\infty} \mathrm{d}\epsilon
			\left( - \frac{\partial f_{\mathrm{FD}}}{\partial \epsilon} \right)
				\sigma (\epsilon)
\label{eq:L_11}
,\end{align}
and the longitudinal charge current response coefficient to the temperature gradient, which is called thermo-electric conductivity, is calculated as
\begin{align}
L_{12}
	& = \frac{1}{- e} \int_{-\infty}^{\infty} \mathrm{d}\epsilon
			\left( - \frac{\partial f_{\mathrm{FD}}}{\partial \epsilon} \right) (\epsilon - \mu)
				\sigma (\epsilon)
\label{eq:L_12}
,\end{align}
where $f_{\mathrm{FD}} (\epsilon) = \{ e^{(\epsilon - \mu)/\kB T} + 1 \}^{-1}$ is the Fermi-Dirac distribution function, and
\begin{align}
\sigma (\epsilon)
	& = \frac{e^2 v^2 \tau (\epsilon)}{2} \frac{(\epsilon^2 - \Delta^2)(\epsilon^2 + \Delta^2)}{\epsilon^2 (\epsilon^2 + 2 \Delta^2)}
		\nu (\epsilon)
.\end{align}
$\tau (\epsilon)$ is the lifetime of electron, and $\nu (\epsilon)$ is DOS given by Eq.~(\ref{eq:DOS}).
Note that Eq.~(\ref{eq:L_11}) reads $L_{11} = \sigma (\eF)$ at $T = 0$, which means that $\sigma (\epsilon)$ describes the electric conductivity at zero temperature.


At the low temperature, we use the Sommerfeld expansion
\begin{align}
\int_{-\infty}^{\infty} \mathrm{d}\epsilon H (\epsilon) \left( - \frac{\partial f_{\mathrm{FD}}}{\partial \epsilon} \right)
	& = H (\mu)
		+\frac{\pi^2}{6} H'' (\mu) (\kB T)^2
		+ \cdots
,\end{align}
which leads to Mott's formula
\begin{align}
L_{12}
	& = \frac{\pi^2}{3} (\kB T)^2
		\frac{\partial}{\partial \eF} \left( \frac{1}{- e} \sigma (\eF) \right)
\label{eq:L12_low}
,\end{align}
where $\eF$ is the Fermi energy ($\mu$ is equivalent to the Fermi energy at absolute zero).
Hence, the Seebeck coefficient is shown as
\begin{align}
S
	& = \frac{\pi^2}{3} \frac{\kB^2 T}{- e}
		\frac{\partial \ln \sigma (\eF)}{\partial \eF}
\label{eq:S_low}
.\end{align}


Figure~\ref{fig:vs_mu} depicts the chemical potential dependences of the response coefficients $L_{11}$, $L_{12}$, and the Seebeck coefficient $S$.
The electric conductivity $L_{11}$ does not change so much at the lower temperature ($\kB T / \Delta \lesssim 0.1$) as shown in Fig.~\ref{fig:vs_mu}~(a), while the thermo-electric conductivity $L_{12}$ has the relatively large dependence on the temperature [Fig.~\ref{fig:vs_mu}~(b)], so that the Seebeck coefficient defined by Eq.~(\ref{eq:Seebeck}) has a peak structure~[Fig.~\ref{fig:vs_mu}~(c)].
We show the chemical potential dependences in a wider range of the temperature in Figs.~\ref{fig:vs_mu}~(d)--(f).
For $\kB T / \Delta \gtrsim 1$, the conductivities $L_{11}$ and $L_{12}$ have finite values even in the band gap $|\mu| / \Delta < 1$ [Figs.~\ref{fig:vs_mu}~(d) and (e)].
This behavior means that the electron and hole pairs are excited by thermal fluctuation, and they contribute to the conductivities, which is well-known in semiconductor physics and superconducting junctions~\cite{tinkham2004}.
As for the Seebeck coefficient, the peak structure becomes broader as the temperature arises, and the structure disappears for $\kB T / \Delta \gtrsim 1$ as shown in Fig.~\ref{fig:vs_mu}~(f).
We discuss the peak structure using a phenomenological analysis in semiconductors after showing the temperature dependences.

Here, we evaluate the Seebeck coefficient $S$ by assuming the rigid band model for doped bismuth.
Assume that one can prepare an appropriate doping, where the chemical potential becomes $\mu \simeq 0.1 \Delta$ that lies in the band gap.
For such a case, the Seebeck coefficient takes $S \simeq - 20 \,\kB / e \simeq - 1.7 \, \mathrm{m eV / K}$ at $\kB T / \Delta = 0.05$ with $\Delta = 15 \,\mathrm{m e V}$, which corresponds to $T = 8.7 \,\mathrm{K}$.
Although the contribution to the Seebeck effect from the holes in the $T$ point should be taken into account, it would not be significant because it should be proportional to $T$ at low temperatures.

\begin{figure*}[t]
\centering
\includegraphics[width=0.8\linewidth]{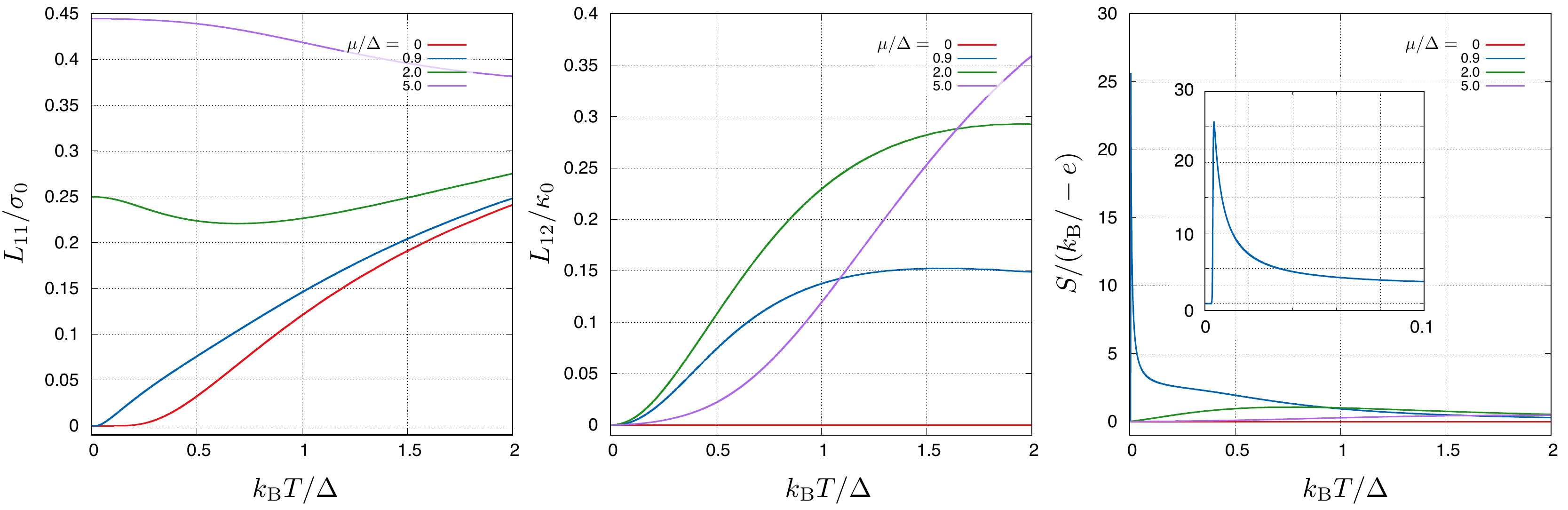}
\caption{\label{fig:vs_T}%
Temperature dependences of the response coefficients $L_{11}$, $L_{12}$, and the Seebeck coefficient $S$ at various chemical potentials.
The units are given in the caption of Fig.~\ref{fig:vs_mu}.
At $\mu = 0$, the response coefficient $L_{12}$ is zero, and hence the Seebeck coefficient $S$ is also zero.
The response coefficient $L_{12}$ and the Seebeck coefficient $S$ become zero when $T \to 0$ for all the chemical potentials.
}
\end{figure*}
Figure~\ref{fig:vs_T} shows the temperature dependence of the response coefficients $L_{11}$, $L_{12}$, and the Seebeck coefficient $S$.
The electric conductivity $L_{11}$ for $\mu / \Delta = 0$ and $0.9$ are zero for sufficiently low temperature since the system has no Fermi surface, while $L_{11}$ for $\mu / \Delta = 2$ and $5$ are metallic even in the absolute zero, as shown in Fig.~\ref{fig:vs_T}~(a).
The thermo-electric conductivity $L_{12}$ and the Seebeck coefficient $S$ become zero when $T \to 0$ for all the chemical potentials [Fig.~\ref{fig:vs_T}~(b) and (c)].
However, the Seebeck coefficient, when the chemical potential lies in the band gap (and $\mu \neq 0$), has different dependence on the temperature from those when the chemical potential lies out of the band gap, and has a peak structure [see the blue line ($\mu / \Delta = 0.9$) in the inset of Fig.~\ref{fig:vs_T}~(c)].
We discuss the peak structure using the phenomenological analysis shortly.
Note that the metallic case ($\mu / \Delta > 1$) have no surprising dependence of $S$ on temperature, and $L_{12}$ and $S$ in this case can be approximately described by Eqs.~(\ref{eq:L12_low}) and (\ref{eq:S_low}), respectively.

\begin{figure}[htpb]
\centering
\includegraphics[width=0.7\linewidth]{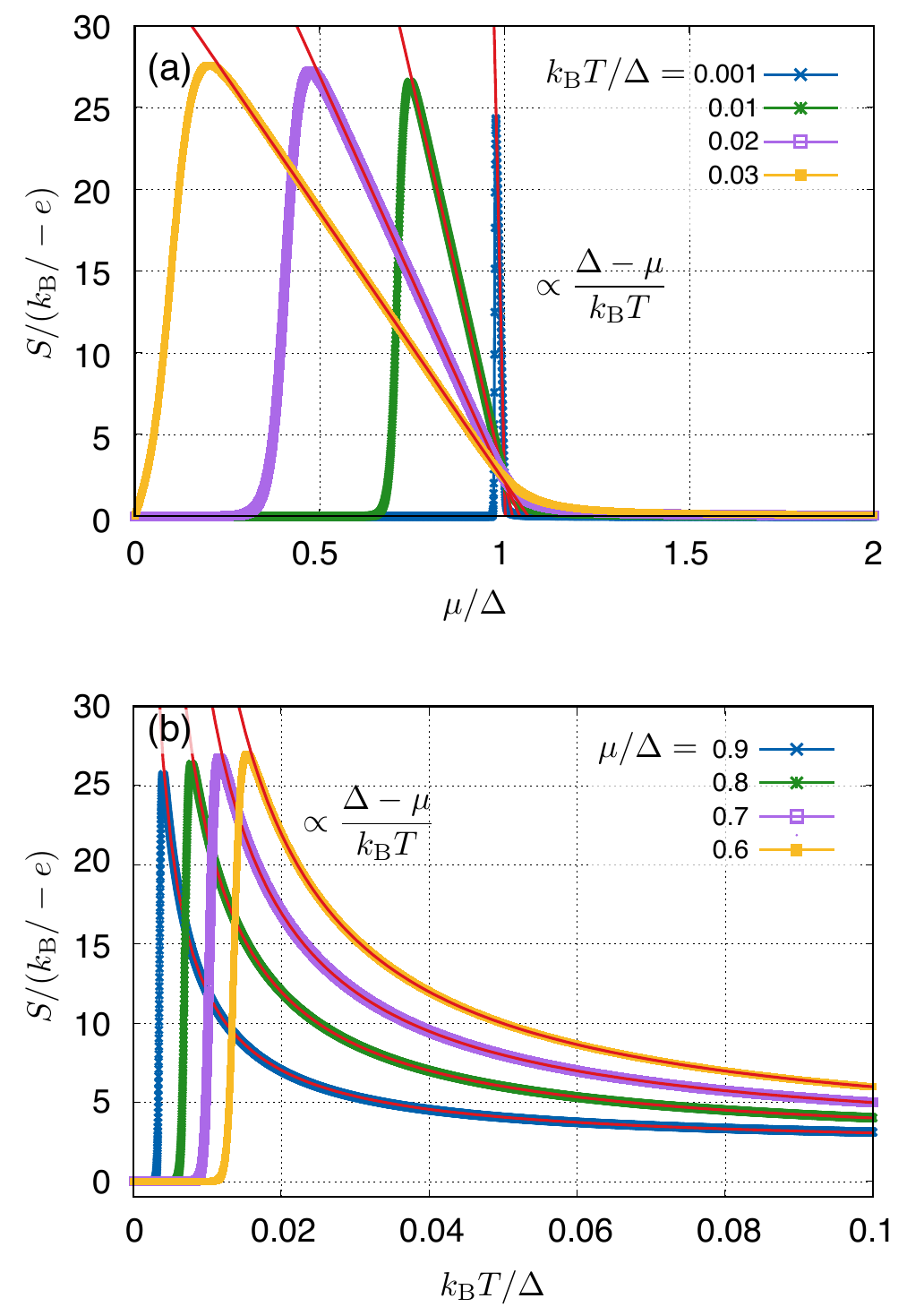}
\caption{\label{fig:Seebeck_fit}
Fittings of the Seebeck coefficient $S$ by the phenomenological function $S_{\mathrm{semi}}$ (a)~for the chemical potential dependence and (b)~for the temperature dependence.
}
\end{figure}
Now, we discuss the peak structures in the chemical potential and temperature dependences of the Seebeck coefficient.
These peak structures appear when the chemical potential lies in the band gap [Figs.~\ref{fig:vs_mu}~(c) and \ref{fig:vs_T}~(c)].
Hence, we regards the system as a semiconductor and apply the phenomenological theory for semiconductors~\cite{behnia2015,mott2012}.
The theory says that $L_{11} \propto \exp[ (\Delta - \mu) / \kB T]$ and $- e L_{12} \propto (\Delta - \mu) \exp[ (\Delta - \mu) / \kB T]$, which yields
\begin{align}
S_{\mathrm{semi}}
	& = \frac{\kB}{-e} \frac{\Delta - \mu}{\kB T}
.\end{align}
By using $S_{\mathrm{semi}}$ for fitting the $\mu$- and $T$-dependences of the Seebeck coefficient, we obtain Fig.~\ref{fig:Seebeck_fit}~(a) and (b), respectively.
The red lines describing $S_{\mathrm{semi}}$ are in good agreements with the $\mu$- and $T$-dependences of $S$.
These agreements indicate that the Seebeck effect in the 3D Dirac electron system when the chemical potential lies in the band gap has the similar behavior to that in semiconductors.
This is mainly because the Seebeck effect is related to the charge degree of freedom and not to the spin degree of freedom.
It would be different from usual semiconductors for the phenomena related to spin, such as the spin Nernst effect.
Note that we can see the deviations for the much lower $\mu / \Delta$ and $\kB T / \Delta$, but we do not know the mechanism yet.

As discussed briefly in Ref.~\onlinecite{fukazawa2017}, we discuss the case of another type of impurity potential which is proportional to $V' \propto \rho_0 + \rho_1$ obtained from the $\bk \cdot \bp$ theory~\cite{sakai1981}, while we have assumed that the impurity potential is proportional to $V \propto \rho_0$.
The self energy for $V'$ contains the only $\gamma_0$-term.
Moreover, the ladder-type vertex correction for $V'$ is found to be $\mathcal{O} (n_{\mathrm{i}})$.
However, Eqs.~(\ref{eq:L_11}) and (\ref{eq:L_12}) do not change even in the case of $V'$, while $\sigma (\epsilon)$ changes that replacing $U / (1 - U)$ with $U$.
Hence, only the quantitative difference in $L_{11}$ and $L_{12}$ arises between the cases of $V$ and $V'$.
The Seebeck coefficient is more robust when changing the type of impurity potential than $L_{11}$ and $L_{12}$, since the ratio only contributes to the Seebeck coefficient.

We compare the results with the experiments~\cite{chandrasekhar1959,das1987}.
The Seebeck coefficient in experiments are in the order of $- 10 \,\mu \mathrm{V / K}$ near the room temperature, and our results indicate that the theoretical value is in the order of $- \kB / e \simeq - 8.6 \, \mu \mathrm{V / K}$, which is in relatively good agreement with the experiments, while we does not consider the contribution from the holes near the $T$ point.
Note that, since our model is isotropic, we cannot discuss the anisotropy of the Seebeck coefficient, which is thus a future work.
Since the experimental data are near the room temperature, effects of phonon may be included, although we do not consider them.
Hence, for quantitative comparison with the experiments, the calculation of effects due to phonon may be needed, which is also a future work.

\section{Conclusion}
We have considered the 3D Dirac electron system and calculated the Seebeck effect based on the linear response theory with Luttinger's gravitational potential.
The Seebeck coefficient $S$ is defined using the longitudinal responses of the charge current to the electric field and to the temperature gradient, whose response coefficients are $L_{11}$ and $L_{12}$, respectively.
We confirm that $L_{11}$ and $L_{12}$ for low temperatures are related through Mott's formula.
We discuss the dependences of $L_{11}$, $L_{12}$, and $S$ on the chemical potential $\mu$ and the temperature $T$.
The Seebeck coefficient $S$, when $|\mu| < \Delta$ with the band gap $2 \Delta$, can be fitted by the function $S_{\mathrm{semi}} \propto (\Delta - \mu) / \kB T$ derived from the phenomenological theory for semiconductors.
We also evaluate the Seebeck coefficient by assuming the doped bismuth, which is the relatively large value $S \simeq 1.7 \,\mathrm{m V/K}$ at $T \simeq 8.7 \,\mathrm{K}$.

\begin{acknowledgments}
This work was supported by Grants-in-Aid for Scientific Research from the Japan Society for the Promotion of Science (Grant No. JP18H01162), and by JST-Mirai Program (Grant No. JPMJMI19A1).  
\end{acknowledgments}

\appendix
\section{\label{apx:heat_current}Heat current operator}
In this Appendix, we derive the heat current operator in the Dirac system.
Consider the system given by
\begin{align}
\tilde{\mathscr{H}}
	& = \mathscr{H} - \mu \mathscr{N}
\notag \\ 
	& = \int \mathrm{d}\br \psi^{\dagger} (\br) \left(
		- v \rho_2 \bp \cdot \bm{\sigma}
		+ \Delta \rho_3
		- \mu \rho_0
	\right) \psi (\br)
\\
	& = \sum_{\bk} c_{\bk}^{\dagger} \left(
		- \hbar v \rho_2 \bk \cdot \bm{\sigma}
		+ \Delta \rho_3
		- \mu \rho_0
	\right) c_{\bk}^{}
\label{eq:tilde_H}
,\end{align}
where $\psi^{(\dagger)} (\br)$ is the field operator of the electron, $\mu$ is the chemical potential, and $\mathscr{N}$ is the electron number.
We make the Hamiltonian symmetric as
\begin{align}
\tilde{\mathscr{H}}
	& = \int \mathrm{d}\br Q (\br)
\end{align}
with the heat operator
\begin{align}
Q (\br)
	& = \psi^{\dagger} (\br) \left(
		\frac{- \hbar v}{2 \zi} \rho_2 \bm{\sigma} \cdot \left( - \overleftarrow{\bm{\nabla}} + \overrightarrow{\bm{\nabla}}  \right)
		+ \Delta \rho_3
		- \mu \rho_0
	\right) \psi (\br)
\notag \\
	& \equiv \psi^{\dagger} (\br) \bar{\mathcal{H}} \psi (\br)
.\end{align}

The continuity equation for the heat operator is calculated as
\begin{align}
\frac{\mathrm{d} Q (\br)}{\mathrm{d} t}
	& = \dot{\psi}^{\dagger} \bar{\mathcal{H}} \psi + \psi^{\dagger} \bar{\mathcal{H}} \dot{\psi}
\notag \\
	& = - \bm{\nabla} \cdot \left\{
		- \frac{\hbar v}{2 \zi} \left( \dot{\psi}^{\dagger} \rho_2 \bm{\sigma} \psi - \psi^{\dagger} \rho_2 \bm{\sigma} \dot{\psi} \right)
	\right\}
\notag \\ & \hspace{1em}
	+ \dot{\psi}^{\dagger} \left(
		- \frac{\hbar v}{\zi} \rho_2 \bm{\sigma} \cdot \overrightarrow{\bm{\nabla}}
		+ \Delta \rho_3
		- \mu \rho_0
	\right) \psi
\notag \\ & \hspace{1em}
	+ \psi^{\dagger} \left(
		+ \frac{\hbar v}{\zi} \rho_2 \bm{\sigma} \cdot \overleftarrow{\bm{\nabla}}
		+ \Delta \rho_3
		- \mu \rho_0
	\right) \dot{\psi}
.\end{align}
The last two terms cancel out since the following relation
\begin{subequations}
\begin{align}
\zi \hbar \frac{\mathrm{d} \psi (\br)}{\mathrm{d} t}
	& = \left(
		- \frac{\hbar v}{\zi} \rho_2 \bm{\sigma} \cdot \overrightarrow{\bm{\nabla}}
		+ \Delta \rho_3
		- \mu \rho_0
	\right) \psi (\br)
, \\
- \zi \hbar \frac{\mathrm{d} \psi^{\dagger} (\br)}{\mathrm{d} t}
	& = \psi^{\dagger} (\br) \left(
		+ \frac{\hbar v}{\zi} \rho_2 \bm{\sigma} \cdot \overleftarrow{\bm{\nabla}}
		+ \Delta \rho_3
		- \mu \rho_0
	\right)
.\end{align}
\label{eqs:EOM}
\end{subequations}
Hence, the continuity equation is obtained as $\mathrm{d} Q (\br) / \mathrm{d} t = - \bm{\nabla} \cdot \bm{j}_{\mathrm{Q}} (\br)$ with the heat current operator
\begin{align}
\bm{j}_{\mathrm{Q}} (\br)
	& = - \frac{\hbar v}{2 \zi} \left( \dot{\psi}^{\dagger} \rho_2 \bm{\sigma} \psi - \psi^{\dagger} \rho_2 \bm{\sigma} \dot{\psi} \right)
\end{align}
in real time domain.
Using $t = - \zi \hbar \tau$, we obtain the heat current operator in the imaginary time domain as
\begin{align}
\bm{j}_{\mathrm{Q}} (\br)
	& = - \frac{v}{2} \left( \dot{\psi}^{\dagger} \rho_2 \bm{\sigma} \psi - \psi^{\dagger} \rho_2 \bm{\sigma} \dot{\psi} \right)
.\end{align}

\section{\label{apx:relation}Derivation of Eq.~(\ref{eqs:derivatives})}
In this Appendix, we derive the relation~(\ref{eqs:derivatives}).
First, we see the $\tau$-derivative of the thermal Green's function $\mathcal{G}_{\bk} (\tau - \tau') = - \langle \mathrm{T}_{\tau} c_{\bk}^{} (\tau) c_{\bk}^{\dagger} (\tau') \rangle$.
Using the definition of the time-ordering operator in the thermal Green's function, we have
\begin{align}
\mathcal{G}_{\bk} (\tau - \tau')
	& = - \Theta (\tau - \tau') \langle c_{\bk}^{} (\tau) c_{\bk}^{\dagger} (\tau') \rangle
\notag \\ & \hspace{1em}
		+ \Theta (\tau' - \tau) \langle c_{\bk}^{\dagger} (\tau') c_{\bk}^{} (\tau) \rangle
\label{eq:Gk_apx}
,\end{align}
and its $\tau$-derivation is obtained as
\begin{align}
\frac{\partial}{\partial \tau} \mathcal{G}_{\bk} (\tau - \tau')
	& = - \delta (\tau - \tau') \langle c_{\bk}^{} (\tau) c_{\bk}^{\dagger} (\tau') \rangle
\notag \\ & \hspace{1em}
		- \delta (\tau' - \tau) \langle c_{\bk}^{\dagger} (\tau') c_{\bk}^{} (\tau) \rangle
\notag \\ & \hspace{1em}
		- \left\langle \mathrm{T}_{\tau} \frac{\partial c_{\bk}^{} (\tau)}{\partial \tau} c_{\bk}^{\dagger} (\tau') \right\rangle
.\end{align}
The first two terms in the right hand side read
\begin{align}
& - \delta (\tau - \tau') \langle c_{\bk}^{} (\tau) c_{\bk}^{\dagger} (\tau') \rangle
- \delta (\tau' - \tau) \langle c_{\bk}^{\dagger} (\tau') c_{\bk}^{} (\tau) \rangle
\notag \\
	& = - \delta (\tau - \tau')
\end{align}
from the anticommutator $\langle c_{\bk}^{} (\tau) c_{\bk}^{\dagger} (\tau) + c_{\bk}^{\dagger} (\tau) c_{\bk}^{} (\tau) \rangle = 1$.
Hence,
\begin{align}
\frac{\partial}{\partial \tau} \mathcal{G}_{\bk} (\tau - \tau')
	& = - \delta (\tau - \tau')
		- \left\langle \mathrm{T}_{\tau} \dot{c}_{\bk}^{} (\tau) c_{\bk}^{\dagger} (\tau') \right\rangle
\label{eq:eom_G}
.\end{align}
Then, we see the Heisenberg equation for the field operator $c_{\bk} (\tau)$, which is given as
\begin{align}
\dot{c}_{\bk}^{} (\tau)
	& = e^{\tau \tilde{\mathscr{H}}} (\tilde{\mathscr{H}} c_{\bk}^{} - c_{\bk}^{} \tilde{\mathscr{H}}) e^{- \tau \tilde{\mathscr{H}}}
\notag \\
	& = e^{\tau \tilde{\mathscr{H}}} \dot{c}_{\bk} e^{- \tau \tilde{\mathscr{H}}}
\label{eq:eom_c}
,\end{align}
where $\tilde{\mathscr{H}}$ is defined by Eq.~(\ref{eq:tilde_H}).
Substituting Eq.~(\ref{eq:eom_c}) into Eq.~(\ref{eq:eom_G}), we have
\begin{align}
\frac{\partial}{\partial \tau} \mathcal{G}_{\bk} (\tau - \tau')
	& = - \delta (\tau - \tau')
		- \left\langle \mathrm{T}_{\tau} \dot{c}_{\bk}^{} c_{\bk}^{\dagger} (\tau' - \tau) \right\rangle
,\end{align}
which leads to Eq.~(\ref{eq:derivatives-b}) by replacing $\tau' - \tau$ with $\tau$.

Similarly, taking the $\tau'$-derivative for Eq.~(\ref{eq:Gk_apx}), we have
\begin{align}
\frac{\partial}{\partial \tau'} \mathcal{G}_{\bk} (\tau - \tau')
	& = \delta (\tau - \tau') \langle c_{\bk}^{} (\tau) c_{\bk}^{\dagger} (\tau') \rangle
\notag \\ & \hspace{1em}
		+ \delta (\tau' - \tau) \langle c_{\bk}^{\dagger} (\tau') c_{\bk}^{} (\tau) \rangle
\notag \\ & \hspace{1em}
		- \left\langle \mathrm{T}_{\tau} c_{\bk}^{} (\tau) \frac{\partial c_{\bk}^{\dagger} (\tau')}{\partial \tau'}  \right\rangle
\notag \\
	& = \delta (\tau - \tau')
		- \left\langle \mathrm{T}_{\tau} c_{\bk}^{} (\tau - \tau') \dot{c}_{\bk}^{\dagger} \right\rangle
,\end{align}
which leads to Eq.~(\ref{eq:derivatives-a}) by replacing $\tau - \tau'$ with $\tau$.

\onecolumngrid
\section{\label{apx:detail:correlation}Calculation detail of correlation functions}
Here, we show the detail of the calculation procedures for obtaining Eqs.~(\ref{eq:L_11-tochu}) and (\ref{eq:L_12-tochu}).
Using the thermal Green's function and taking the average on the impurity positions, Eq.~(\ref{eq:chi_c}) is rewritten as
\begin{align}
\chi_c (\zi \omega_{\lambda})
	& = - \frac{e^2 v^2}{\beta \Omega} \sum_n \sum_{\bk}
	\tr \left[
		\rho_2 \sigma^x \mathcal{G}_{\bk} (\zi \epsilon_n^{+})
		\Lambda_{2,x} (\zi \epsilon_n^{+}, \zi \epsilon_n) \mathcal{G}_{\bk} (\zi \epsilon_n)
	\right]
,\end{align}
where $\zi \epsilon_n^{+} = \zi \epsilon_n + \zi \omega_{\lambda}$, and $\Lambda_{2,x} (\zi \epsilon_n^{+}, \zi \epsilon_n)$ is the full velocity vertex including the ladder-type vertex corrections, which is defined by
\begin{align}
\Lambda_{2,x} (\zi \epsilon_n^{+}, \zi \epsilon_n)
	& = \rho_2 \sigma^x
		+ \frac{n_{\mathrm{i}} u^2}{\Omega} \sum_{\bk}
			\mathcal{G}_{\bk} (\zi \epsilon_n^{+})
			\Lambda_{2,x} (\zi \epsilon_n^{+}, \zi \epsilon_n)
			\mathcal{G}_{\bk} (\zi \epsilon_n)
.\end{align}

The correlation function $\chi_Q (\zi \omega_{\lambda})$ is also rewritten as follows.
For the case without the ladder-type vertex corrections, it reads
\begin{align}
\chi_Q (\zi \omega_{\lambda})
	& = - \frac{e v^2}{2 \Omega} \int_0^{\beta} \mathrm{d}\tau e^{\zi \omega_{\lambda} \tau} \sum_{\bk}
	 \tr \left[
		\rho_2 \sigma^x \langle \mathrm{T}_{\tau} c_{\bk} (\tau) \dot{c}_{\bk}^{\dagger} \rangle
		\rho_2 \sigma^x \mathcal{G}_{\bk} (- \tau)
		- \rho_2 \sigma^x \mathcal{G}_{\bk} (\tau)
		\rho_2 \sigma^x \langle \mathrm{T}_{\tau} \dot{c}_{\bk} c_{\bk}^{\dagger} (\tau) \rangle
	\right]
,\end{align}
and using the relation~(\ref{eqs:derivatives}), we have
\begin{align}
\chi_Q (\zi \omega_{\lambda})
	& = - \frac{e v^2}{2 \Omega} \int_0^{\beta} \mathrm{d}\tau e^{\zi \omega_{\lambda} \tau} \sum_{\bk}
	\tr \left[
		\rho_2 \sigma^x \left( \frac{\mathrm{d}}{\mathrm{d} \tau} \mathcal{G}_{\bk} (\tau) \right)
		\rho_2 \sigma^x \mathcal{G}_{\bk} (- \tau)
		- \rho_2 \sigma^x \mathcal{G}_{\bk} (\tau)
		\rho_2 \sigma^x \left( \frac{\mathrm{d}}{\mathrm{d} \tau} \mathcal{G}_{\bk} (- \tau) \right)
	\right]
\notag \\
	& = \frac{e v^2}{\beta \Omega} \sum_n \sum_{\bk} \tr \biggl[
		\rho_2 \sigma^x \mathcal{G}_{\bk} (\zi \epsilon_n + \zi \omega_{\lambda})
		\left( \zi \epsilon_n + \frac{\zi \omega_{\lambda}}{2} \right) \rho_2 \sigma^x \mathcal{G}_{\bk} (\zi \epsilon_n)
	\biggr]
.\end{align}
Considering the vertex corrections, we have
\begin{align}
\chi_Q (\zi \omega_{\lambda})
	& = \frac{e v^2}{\beta \Omega} \sum_{n} \sum_{\bk} \tr \biggl[
		\rho_2 \sigma^x \mathcal{G}_{\bk} (\zi \epsilon_n^{+})
		\left( \zi \epsilon_n + \frac{\zi \omega_{\lambda}}{2} \right) \Lambda_{2, x} (\zi \epsilon_n^{+}, \zi \epsilon_n)  \mathcal{G}_{\bk} (\zi \epsilon_n)
	\biggr]
.\end{align}

Rewriting the Matsubara summation into the contour integral in $\chi_c (\zi \omega_{\lambda})$ and $\chi_Q (\zi \omega_{\lambda})$, and taking the analytic continuation $\zi \omega_{\lambda} \to \hbar \omega + \zi 0$, we obtain
\begin{align}
\chi_c (\omega)
	& = - \frac{\zi e^2 v^2}{2 \Omega} \sum_{\bk} \int_{-\infty}^{\infty} \frac{\mathrm{d}\epsilon}{2 \pi}
		\biggl\{
			\left( f (\epsilon_{+}) - f (\epsilon_{-}) \right)
			\tr \left[ \rho_2 \sigma^x G^{\R}_{\bk} (\epsilon_{+}) \Lambda_{2, x}^{\R \A} G^{\A}_{\bk} (\epsilon_{-}) \right]
\notag \\ & \hspace{8em}
			+ f (\epsilon_{-}) \tr \left[ \rho_2 \sigma^x G^{\R}_{\bk} (\epsilon_{+}) \Lambda_{2, x}^{\R \R} G^{\R}_{\bk} (\epsilon_{-}) \right]
			- f (\epsilon_{+}) \tr \left[ \rho_2 \sigma^x G^{\A}_{\bk} (\epsilon_{+}) \Lambda_{2, x}^{\A \A} G^{\A}_{\bk} (\epsilon_{-}) \right]
		\biggr\}
\label{eq:chi_c_appendix}
, \\
\chi_Q (\omega)
	& = \frac{\zi e v^2}{2 \Omega} \sum_{\bk} \int_{-\infty}^{\infty} \frac{\mathrm{d}\epsilon}{2 \pi}
		\epsilon \biggl\{
			\left( f (\epsilon_{+}) - f (\epsilon_{-}) \right)
			\tr \left[ \rho_2 \sigma^x G^{\R}_{\bk} (\epsilon_{+}) \Lambda_{2, x}^{\R \A} G^{\A}_{\bk} (\epsilon_{-}) \right]
\notag \\ & \hspace{8em}
			+ f (\epsilon_{-}) \tr \left[ \rho_2 \sigma^x G^{\R}_{\bk} (\epsilon_{+}) \Lambda_{2, x}^{\R \R} G^{\R}_{\bk} (\epsilon_{-}) \right]
			- f (\epsilon_{+}) \tr \left[ \rho_2 \sigma^x G^{\A}_{\bk} (\epsilon_{+}) \Lambda_{2, x}^{\A \A} G^{\A}_{\bk} (\epsilon_{-}) \right]
		\biggr\}
\label{eq:chi_Q_appendix}
,\end{align}
where $\epsilon_{\pm} = \epsilon \pm \hbar \omega / 2$, $f (\epsilon) = (e^{\beta \epsilon} + 1)^{-1}$, $G^{\R}_{\bk} (\epsilon)$ are the retarded Green's function given by Eq.~(\ref{eq:retarded_G_def}), $G^{\A}_{\bk} (\epsilon)$ is the advanced Green's function obtained by replacing $\zi \gamma_0$ and $\zi \gamma_3$ with $- \zi \gamma_0$ and $- \zi \gamma_3$ in $G^{\R}_{\bk} (\epsilon)$, respectively, and the full velocity vertex $\Lambda_{2, x}^{\R \A}$, $\Lambda_{2, x}^{\R \R}$, and $\Lambda_{2, x}^{\A \A}$ are given as
\begin{align}
\Lambda_{2, x}^{\X \Y}
	& = \rho_2 \sigma^x
		+ \frac{n_{\mathrm{i}} u^2}{\Omega} \sum_{\bk} G^{\X}_{\bk} (\epsilon_{+}) \Lambda_{2, x}^{\X \Y} G^{\Y}_{\bk} (\epsilon_{-})
, \qquad
(\X, \Y \in \{\R, \A\})
.\end{align}
Note that $\Lambda_{2, x}^{\R \A}$ in this Appendix is equivalent to $\Lambda_{2, x}$ [Eq.~(\ref{eq:Lambda_2x})] in the main text.
We also note that Eqs.~(\ref{eq:chi_c_appendix}) and (\ref{eq:chi_Q_appendix}) are different only in the factor $- \epsilon / e$.

We extract the $\omega$-linear terms in $\chi_c (\omega)$ and $\chi_Q (\omega)$, which are given as
\begin{align}
L_{11}
	& = \frac{\hbar e^2 v^2}{2 \Omega} \sum_{\bk} \int_{-\infty}^{\infty} \frac{\mathrm{d}\epsilon}{2 \pi}
		\biggl\{
			\left( - \frac{\partial f}{\partial \epsilon} \right) \tr \left[
				\rho_2 \sigma^x G^{\R}_{\bk} (\epsilon) \Lambda_{2, x}^{\R \A} G^{\A}_{\bk} (\epsilon)
				- \re \{ \rho_2 \sigma^x G^{\R}_{\bk} (\epsilon) \Lambda_{2, x}^{\R \R} G^{\R}_{\bk} (\epsilon) \}
			\right]
\notag \\ & \hspace{13em}
			- \zi f (\epsilon) (\partial_{\epsilon} - \partial_{\epsilon'}) \tr \left[
				\im \{ \rho_2 \sigma^x G^{\R}_{\bk} (\epsilon) \Lambda_{2, x}^{\R \R} G^{\R}_{\bk} (\epsilon') \}
			\right] \Big|_{\epsilon' \to \epsilon}
		\biggr\}
, \\
L_{12}
	& = - \frac{\hbar e v^2}{2 \Omega} \sum_{\bk} \int_{-\infty}^{\infty} \frac{\mathrm{d}\epsilon}{2 \pi}
		\epsilon \biggl\{
			\left( - \frac{\partial f}{\partial \epsilon} \right) \tr \left[
				\rho_2 \sigma^x G^{\R}_{\bk} (\epsilon) \Lambda_{2, x}^{\R \A} G^{\A}_{\bk} (\epsilon)
				- \re \{ \rho_2 \sigma^x G^{\R}_{\bk} (\epsilon) \Lambda_{2, x}^{\R \R} G^{\R}_{\bk} (\epsilon) \}
			\right]
\notag \\ & \hspace{13em}
			- \zi f (\epsilon) (\partial_{\epsilon} - \partial_{\epsilon'}) \tr \left[
				\im \{ \rho_2 \sigma^x G^{\R}_{\bk} (\epsilon) \Lambda_{2, x}^{\R \R} G^{\R}_{\bk} (\epsilon') \}
			\right] \Big|_{\epsilon' \to \epsilon}
		\biggr\}
,\end{align}
where $\re [\cdots]$ and $\im [\cdots]$ are defined as
\begin{align}
\re \{ P^{\R} \}
	= \frac{1}{2} \left[ P^{\R} + P^{\A} \right]
, & \qquad
\im \{ P^{\R} \}
	= \frac{1}{2 \zi} \left[ P^{\R} - P^{\A} \right]
.\end{align}
Then, we neglect the terms which include only $G^{\R}_{\bk}$ or $G^{\A}_{\bk}$, since they contribute only in the higher orders with respect to $q = \mu \tau /\hbar$ for $q \gg 1$.
Finally, we have Eqs.~(\ref{eq:L_11-tochu}) and (\ref{eq:L_12-tochu}) with Eq.~(\ref{eq:sigma-tochu}) in the main text.

\section{\label{apx:detail:vc}Calculation detail of vertex corrections}
Here, we show the calculation of the ladder-type vertex corrections.
Equation~(\ref{eq:Lambda_2x}) reads
\begin{align}
\Lambda_{2,x}
	& = \rho_2 \sigma^x
		+ \frac{n_{\mathrm{i}} u^2}{\Omega} \sum_{\bk} G^{\R}_{\bk} (\epsilon) \rho_2 \sigma^x G^{\A}_{\bk} (\epsilon)
		+ \left( \frac{n_{\mathrm{i}} u^2}{\Omega} \right)^2 \sum_{\bk, \bk'} G^{\R}_{\bk} (\epsilon) G^{\R}_{\bk'} (\epsilon) \rho_2 \sigma^x G^{\A}_{\bk'} (\epsilon) G^{\A}_{\bk} (\epsilon)
		+ \cdots
\label{eq:Lambda_2x_apx}
.\end{align}
Firstly, we calculate the second term in the right hand side, which is computed as
\begin{align}
\frac{n_{\mathrm{i}} u^2}{\Omega} \sum_{\bk} G^{\R}_{\bk} (\epsilon) \rho_2 \sigma^x G^{\A}_{\bk} (\epsilon)
	& = U (\epsilon + \mu) \rho_2 \sigma^x
		- V (\epsilon + \mu)  \rho_1 \sigma^x
,\end{align}
where $U (\epsilon)$ and $V (\epsilon)$ are given by
\begin{align}
U (\epsilon)
	& = \frac{1}{4} \frac{n_{\mathrm{i}} u^2}{\Omega} \sum_{\bk} \tr \left[ G^{\R}_{\bk} (\epsilon - \mu) \rho_2 \sigma^x G^{\A}_{\bk} (\epsilon - \mu) \rho_2 \sigma^x \right]
, \\
V (\epsilon)
	& = \frac{1}{4} \frac{n_{\mathrm{i}} u^2}{\Omega} \sum_{\bk} \tr \left[ G^{\R}_{\bk} (\epsilon - \mu) \rho_2 \sigma^x G^{\A}_{\bk} (\epsilon - \mu) \rho_1 \sigma^x \right]
.\end{align}
Using $U (\epsilon)$ and $V (\epsilon)$, the third term in Eq.~(\ref{eq:Lambda_2x_apx}) is calculated as
\begin{align}
& \left( \frac{n_{\mathrm{i}} u^2}{\Omega} \right)^2 \sum_{\bk, \bk'}
	G^{\R}_{\bk} (\epsilon) G^{\R}_{\bk'} (\epsilon) \rho_2 \sigma^x G^{\A}_{\bk'} (\epsilon) G^{\A}_{\bk} (\epsilon)
\notag \\
	& = \frac{n_{\mathrm{i}} u^2}{\Omega} \sum_{\bk} G^{\R}_{\bk} (\epsilon) \left(
			U (\epsilon + \mu) \rho_2 \sigma^x
			- V (\epsilon + \mu)  \rho_1 \sigma^x
		\right) G^{\A}_{\bk} (\epsilon)
\notag \\
	& = U (\epsilon + \mu) \Bigl\{
			U (\epsilon + \mu) \rho_2 \sigma^x
			- V (\epsilon + \mu)  \rho_1 \sigma^x
		\Bigr\}
		- V (\epsilon + \mu)
			\frac{n_{\mathrm{i}} u^2}{\Omega} \sum_{\bk} G^{\R}_{\bk} (\epsilon) \rho_1 \sigma^x G^{\A}_{\bk} (\epsilon)
\notag \\
	& = U (\epsilon + \mu) \Bigl\{
			U (\epsilon + \mu) \rho_2 \sigma^x
			- V (\epsilon + \mu)  \rho_1 \sigma^x
		\Bigr\}
		- V (\epsilon + \mu) \Bigl\{
			2 U (\epsilon + \mu) \rho_2 \sigma^x
			+ V (\epsilon + \mu)  \rho_1 \sigma^x
		\Bigr\}
\notag \\
	& = \begin{pmatrix}
		0
	&	1
	\end{pmatrix}
	\begin{pmatrix}
		2 U (\epsilon + \mu)
	&	V (\epsilon + \mu)
	\\	- V (\epsilon + \mu)
	&	U (\epsilon + \mu)
	\end{pmatrix}^2
	\begin{pmatrix}
		\rho_1 \sigma^x
	\\	\rho_2 \sigma^x
	\end{pmatrix}
.\end{align}
Here, we used
\begin{align}
\frac{n_{\mathrm{i}} u^2}{\Omega} \sum_{\bk} G^{\R}_{\bk} (\epsilon) \rho_1 \sigma^x G^{\A}_{\bk} (\epsilon)
	& = 2 U (\epsilon + \mu) \rho_1 \sigma^x
		+ V (\epsilon + \mu) \rho_2 \sigma^x
.\end{align}
Similarly, the first and second terms in Eq.~(\ref{eq:Lambda_2x_apx}) are also written as
\begin{align}
\rho_2 \sigma^x
	& = \begin{pmatrix}
		0
	&	1
	\end{pmatrix}
	\begin{pmatrix}
		\rho_1 \sigma^x
	\\	\rho_2 \sigma^x
	\end{pmatrix}
, \\
\frac{n_{\mathrm{i}} u^2}{\Omega} \sum_{\bk} G^{\R}_{\bk} (\epsilon) \rho_2 \sigma^x G^{\A}_{\bk} (\epsilon)
	& = \begin{pmatrix}
		0
	&	1
	\end{pmatrix}
	\begin{pmatrix}
		2 U (\epsilon + \mu)
	&	V (\epsilon + \mu)
	\\	- V (\epsilon + \mu)
	&	U (\epsilon + \mu)
	\end{pmatrix}
	\begin{pmatrix}
		\rho_1 \sigma^x
	\\	\rho_2 \sigma^x
	\end{pmatrix}
.\end{align}
From these, we have
\begin{align}
\Lambda_{2, x}
	& = \begin{pmatrix}
		0
	&	1
	\end{pmatrix}
	\left\{
		\begin{pmatrix}
			1
		&	0
		\\	0
		&	1
		\end{pmatrix}
		+ \begin{pmatrix}
			2 U (\epsilon + \mu)
		&	V (\epsilon + \mu)
		\\	- V (\epsilon + \mu)
		&	U (\epsilon + \mu)
		\end{pmatrix}
		+ \begin{pmatrix}
			2 U (\epsilon + \mu)
		&	V (\epsilon + \mu)
		\\	- V (\epsilon + \mu)
		&	U (\epsilon + \mu)
		\end{pmatrix}^2
		+ \cdots
	\right\}
	\begin{pmatrix}
		\rho_1 \sigma^x
	\\	\rho_2 \sigma^x
	\end{pmatrix}
\notag \\
	& = \begin{pmatrix}
		0
	&	1
	\end{pmatrix}
	\left\{
		A^0
		+ A
		+ A^2
		+ \cdots
	\right\}
	\begin{pmatrix}
		\rho_1 \sigma^x
	\\	\rho_2 \sigma^x
	\end{pmatrix}
,\end{align}
where
\begin{align}
A
	& = \begin{pmatrix}
			2 U (\epsilon + \mu)
		&	V (\epsilon + \mu)
		\\	- V (\epsilon + \mu)
		&	U (\epsilon + \mu)
		\end{pmatrix}
.\end{align}
Introducing the matrix $P$ which diagonalize $A$ and the eigenvalue matrix $E = \mathrm{diag} (\lambda_1, \lambda_2)$; $A = P (P^{-1} A P) P^{-1} = P E P^{-1}$, we find
\begin{align}
A^0 + A + A^2 + \cdots
	& = P \left( 1 + E + E^2 + \cdots \right) P^{-1}
	= P \begin{pmatrix}
		\frac{1}{1 - \lambda_1}
	&	0
	\\	0
	&	\frac{1}{1 - \lambda_2}
	\end{pmatrix}
	P^{-1}
.\end{align}
Hence, we obtain
\begin{align}
\Lambda_{2, x}
	& = \begin{pmatrix}
		0
	&	1
	\end{pmatrix}
	P
	\begin{pmatrix}
		\frac{1}{1 - \lambda_1}
	&	0
	\\	0
	&	\frac{1}{1 - \lambda_2}
	\end{pmatrix}
	P^{-1}
	\begin{pmatrix}
		\rho_1 \sigma^x
	\\	\rho_2 \sigma^x
	\end{pmatrix}
.\end{align}

Before showing the forms of $P$ and $E$, we calculate $U (\epsilon)$ and $V (\epsilon)$.
\begin{align}
U (\epsilon)
	& = \frac{n_{\mathrm{i}} u^2}{\Omega} \sum_{\bk} \frac{1}{| D^{\R}_{\bk} (\epsilon) |^2}
		| g^{\R}_{2,x} (\bk) |^2
, \\
V (\epsilon)
	& = \frac{n_{\mathrm{i}} u^2}{\Omega} \sum_{\bk} \frac{1}{| D^{\R}_{\bk} (\epsilon) |^2}
		\left( g^{\R}_3 (\epsilon) g^{\A}_{0} (\epsilon) - g^{\R}_0 (\epsilon) g^{\A}_3 (\epsilon) \right)
.\end{align}
From the symmetry of the system, $| g^{\R}_{2,x} (\bk) |^2 = \hbar^2 v^2 k^2 / 3 = (\epsilon_k^2 - \Delta^2) / 3$.
We exppress $D^{\R}_{\bk} (\epsilon) = D' + \zi D''$ with
\begin{align}
D'
	& = (\epsilon - \epsilon_k) (\epsilon + \epsilon_k) + \mathcal{O} (n_i^2)
, \\
D''
	& = 2 (\epsilon \gamma_0 (\epsilon) + \Delta \gamma_3 (\epsilon))
,\end{align}
and use the following approximations
\begin{align}
\frac{1}{| D^{\R}_{\bk} (\epsilon) |^2}
	& \simeq \frac{\pi}{| D'' |} \delta (D')
\notag \\
	& \simeq \frac{\pi}{4} \frac{1}{| \epsilon \gamma_0 (\epsilon) + \Delta \gamma_3 (\epsilon) |} \frac{1}{|\epsilon|} \sum_{\eta = \pm} \delta (\epsilon - \eta \epsilon_k)
,\end{align}
where, we assumed small $n_{\mathrm{i}}$ and approximated $D'' / ( (D')^2 + (D'')^2)$ by the $\delta$-function in the first line, and dropped $\gamma_0^2$ and $\gamma_3^2$ in the second line.
From these, we have
\begin{align}
U (\epsilon)
	& = \frac{1}{3} \frac{\epsilon^2 - \Delta^2}{\epsilon^2 + \Delta^2} + \mathcal{O} (n_{\mathrm{i}}^2)
, \\
V (\epsilon)
	& = n_{\mathrm{i}} u^2 \frac{\pi \Delta}{\epsilon^2 + \Delta^2} + \mathcal{O} (n_{\mathrm{i}}^3)
.\end{align}

For small $n_{\mathrm{i}}$, which leads to $\{U (\epsilon + \mu) \}^2 - 4 \{ V (\epsilon + \mu) \}^2 \simeq \{U (\epsilon + \mu) \}^2$, the eigenvalue is obtained as $\lambda_1 = 2 U (\epsilon + \mu)$ and $\lambda_2 = U (\epsilon + \mu)$.
The diagonalization matrix $P$ is calculated as
\begin{align}
P
	\simeq \begin{pmatrix}
		U (\epsilon + \mu)
	&	- V (\epsilon + \mu)
	\\	- V (\epsilon + \mu)
	&	U (\epsilon + \mu)
	\end{pmatrix}
, \qquad
P^{-1}
	\simeq \frac{1}{\{U (\epsilon + \mu) \}^2} \begin{pmatrix}
		U (\epsilon + \mu)
	&	V (\epsilon + \mu)
	\\	V (\epsilon + \mu)
	&	U (\epsilon + \mu)
	\end{pmatrix}
.\end{align}
Hence,
\begin{align}
\Lambda_{2, x}
	& \simeq \begin{pmatrix}
		0
	&	1
	\end{pmatrix}
	\begin{pmatrix}
	\displaystyle
		\frac{1}{1 - 2 U (\epsilon + \mu)}
	& \displaystyle
		\frac{1}{\{1 - 2 U (\epsilon + \mu)\} \{ 1 - U (\epsilon + \mu) \}}
	\\[2ex] \displaystyle
		-\frac{1}{\{1 - 2 U (\epsilon + \mu)\} \{ 1 - U (\epsilon + \mu) \}}
	& \displaystyle
		\frac{1}{1 - U (\epsilon + \mu)}
	\end{pmatrix}
	\begin{pmatrix}
		\rho_1 \sigma^x
	\\	\rho_2 \sigma^x
	\end{pmatrix}
\notag \\
	& = \frac{1}{1 - U (\epsilon + \mu)} \rho_2 \sigma^x
.\end{align}

\bibliography{manuscript}
\end{document}